\documentclass[conference]{IEEEtran}
\usepackage{amsfonts,amsmath}
\usepackage{amssymb, amsmath, graphicx, epsfig, cite,subfigure}

 \newtheorem{theorem}{Theorem}[section]

 \newtheorem{remark}[theorem]{Remark}

\title{Clustering Consumption in Queues: A Scalable Model for Electric Vehicle Scheduling}

\author{Mahnoosh Alizadeh\IEEEauthorrefmark{1}, George Kesidis\IEEEauthorrefmark{2}, and Anna Scaglione\IEEEauthorrefmark{1}\\
\IEEEauthorrefmark{1}University of California Davis~~~ \IEEEauthorrefmark{2}Pennsylvania State University
}

\begin{document}
\maketitle

\begin{abstract}

In this paper, we introduce a scalable model for
the aggregate electricity demand of a fleet of electric vehicles, which can provide the right balance between model simplicity and accuracy. The model is based on classification of tasks with similar energy consumption characteristics into a finite number of clusters. The aggregator responsible for scheduling the charge of the vehicles has two goals: 1) to provide a hard QoS guarantee to the vehicles at the lowest possible cost; 2) to offer load or generation following services to the wholesale market. In order to achieve these goals, we combine the scalable demand model we propose with two scheduling mechanisms, a near-optimal and a heuristic technique. The performance of the two mechanisms is compared under a realistic setting in our numerical experiments.

\end{abstract}

\section{Introduction}\label{sec.intro}
Today, electric power is traded on an hourly basis. However, demand is random in nature and varies within the hour. Since physical constraints of the grid dictate that demand and supply should be instantaneously balanced, a separate market environment exists where fast-ramping generators offer to follow the intra-hour variations of demand in real-time. The service is referred to as {\it load following} and is dispatched by the {\it Independent System Operator (ISO)} \cite{iso}. Consequently, the availability of {\it enough} fast ramping controllable generation is essential to keep the grid safe.  With a widespread integration of renewables, generation quantities will also become random and uncontrollable, and the need for real-time balancing capacity will increase. With the high costs and scarcity of fast ramping generators, this brute-force solution of tweaking the generation to ensure  balance is no longer viable. 

A positive development that could reverse this trend is the electrification of transportation. In fact, with a large scale integration of electric vehicles (EVs) and Plug-in Hybrid Electric Vehicles (PHEV), a large fleet of batteries can act as an invaluable resource for balancing demand/supply at shorter time scales. In this paper, we discuss  a scalable model to incorporate this flexibility as a {\it load} or {\it wind} following reserve.  We assume that an {\it aggregator} is  in charge of scheduling the charge of EVs in real-time to execute intra-hour balancing instructions from the ISO by tweaking its aggregate load.

\subsection{Possible  Scenarios Regarding Charge Flexibility}\label{sec123}
There are three different charge flexibility scenarios for the EVs that differ in terms of the required physical infrastructure:
\begin{enumerate}
\item  Deferrable but non-interruptible charge, with an uncontrollable charging rate (instantaneous power); 
\item Charge can be interrupted at an intermediate state and resumed later. Charging rate is still uncontrollable; 
\item Charging rate is considered controllable. Vehicle-to-grid (V2G) can also be considered in this scenario.
\end{enumerate}
Almost all the previous work is focused on the third scenario, e.g. \cite{markel,donadee, sharkawi,hanhan,tuff,matta, ota}. However, even with the right  infrastructure in place for the instantaneous charging rates to be fully controllable, centralized hard real-time scheduling results in computationally intensive optimizations and is intractable at large scale. As a result, the authors in \cite{galus,subram} have proposed heuristic schemes for  the third scenario. We think that an important first step in designing a {\it scalable} central real-time scheduler is the right form of {\it approximation} in describing the workload/consumption of different tasks.  
 
The most popular approximation in the Smart Grid literature is to ignore the shape of the consumption profiles of different appliances, and model any appliance as a battery with a controllable charging rate, which can be unrealistic. To overcome this issue, our work maps different requests into clusters of demand behavior. Based on the ideas behind quantization and classification, each cluster has an associated {\it code} that approximates the behavior of all requests within the cluster. This code is a scalable multiple description model for reconfiguring the charge of EVs.  The number of clusters determines the level of reconstruction error in the load.

Here, we focus on the second EV charge scenario to present an example of our proposed classification based aggregate load modeling princinples. Naturally, the second scenario requires a much lower infrastructure cost than the third scenario but, it can only deliver a comparable service to the third scenario with a very large population of participants \cite{demanddispatch}. 
Note that the ideas we propose for load classification fit the first and third scenarios as well, as will be seen in future work.

\subsection{The Aggregator's Objective}\label{aggobj}

The aggregator runs a program that offers EV owners cheap tariffs in return for directly managing their charge schedule. We assume that all participating EVs are plugged in before a certain deadline, denoted by $t=0$. No charging requests are accepted after  this time, e.g., after 9 pm. Each customer can choose an individual charging deadline.  We provide a hard quality of service (QoS) guarantee, which ensures that all EVs  will be fully charged by their deadline. We assume that this is always feasible due to the long duration of the night.

Due to the structure of the electricity market, a retailer wishes to buy {\it exactly enough} power to serve its load, {\it no more no less}. This power has to be purchased at least an hour ahead of real-time, based on a {\it prediction} of demand. Real-time deviations from this bulk purchase can be very highly priced. Thus, our aggregator has two operational goals: 1) purchase exactly enough bulk power from the market to serve the EVs with hard QoS. This power profile is constant during hourly intervals. At the beginning of every hour, the aggregator has an opportunity to update its bulk purchase for the next hour. We denote the bulk purchase for a future time $t$, updated at hour $k$, by $P^{(k)}(t)$; 2) to follow real-time load following instructions from the ISO. These instructions, denoted by $a(t)$, are received on a 5-minute time resolution and in the form of a deviation from the bulk hourly purchase $P^{(k)}(t)$. The aggregator is  penalized if its  load deviates from $P^{(k)}(t)+ a(t)$. Thus, the aggregator repeats the following steps in a loop:
\begin{enumerate}
\item Update its bulk purchase $P^{(k)}(t)$ at every hour $k$ to meet hard QoS constraints at lowest possible market cost;
\item Schedule EVs to follow intra-hourly load following signals and adjust consumption to be close to $P^{(k)}(t) + a(t)$. 
\end{enumerate}
Due to space limitations, we mostly focus on the second step and leave the detailed discussion of the first as future work.


\section{Scalable Load Model}\label{sec.g}
We assume that load following dispatch signals $a(t)$ are sent to the aggregator at discrete epochs every 5 minutes, denoted by $t= 1,\ldots,T$. The aggregator wishes to schedule its total load $L(t)$ to be as close as possible to $P^{(k)}(t) + a(t)$. We choose to limit the times at which the aggregator takes action to these epochs. Next, we present the details of our scalable model for $L(t)$, to be used later to design a scheduling strategy.

\begin{figure}
\centering
\includegraphics[width = 0.95\linewidth]{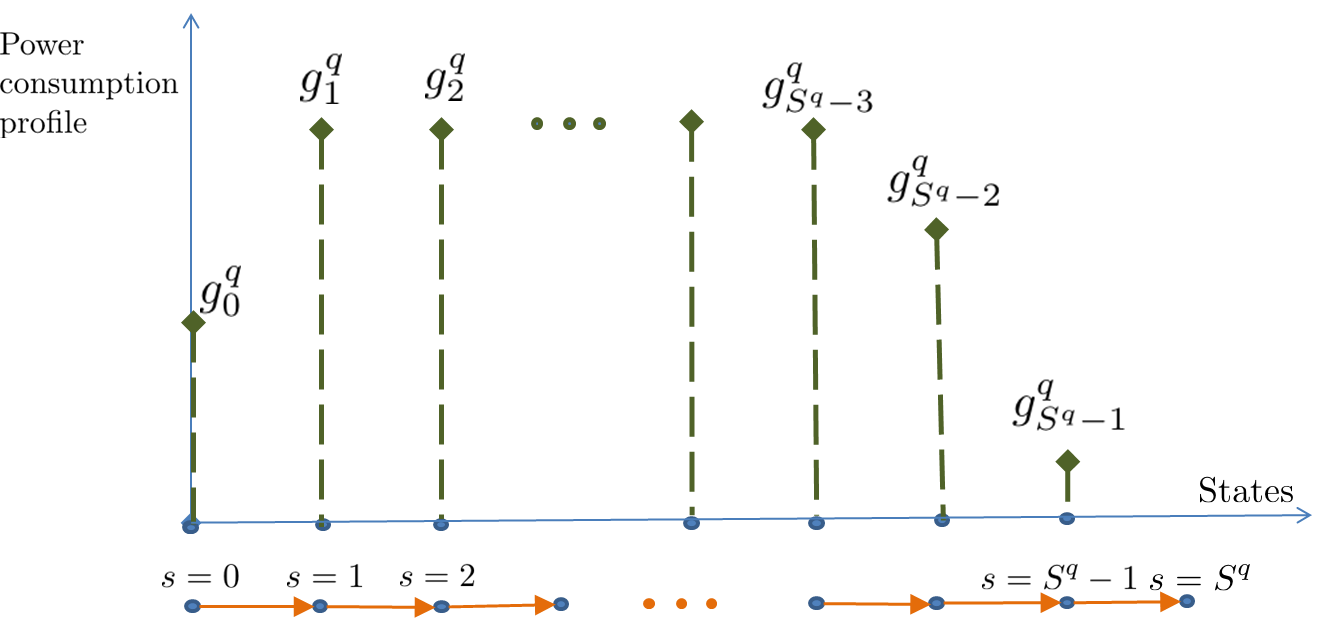}
\label{reg}
\caption{EVs go from one state to another until they receive full charge (reach state $s=S^q$)}
\end{figure}

\subsection{Classification of Charge Requests}
We assume that the set of acceptable power consumption patterns for each EV is characterized by a set of parameters that depend on the physical characteristics of the vehicle, the specifications of the charging station, and the preferences of the customer. These parameters can include, but are not limited to: initial state of charge (SoC), charge deadline, charge rate,  battery capacity, make and model, etc.  For reasons that will be subsequently clear, we separately denote the SoC by a scalar $c$ and bundle the rest of the parameters in a vector $\mathbf{v}$, which we refer to as the {\it characteristic vector}. Consequently, each charge request is fully described by a tuple $(\mathbf{v},c)$.

Notice that the tuple $(\mathbf{v},c)$  can take a very large if not infinite number of values. Thus, the scheduler will  face high computational and data storage costs to keep track of every single EVs characteristics. An important measure taken in this work to alleviate this issue is to classify loads that share similar parameters in a finite number of clusters denoted by a tuple of indices $(q,s)$. The classifier is a mapping
\begin{equation}
 (\mathbf{v},c) \xrightarrow[\mathcal{Q}]{\mbox{Cluster}} (\mathbf{v}_q,c^q_s) \xrightarrow[\mathcal{I}]{\mbox{Cluster indices}} (q,s).
\end{equation}
Notice that the quantization points of the SoC parameter $c$, which we denote by $c_1^q,\ldots,c_{S^q}^q$, depend on all the other parameters bundled in the characteristic vector $\mathbf{v}_q$. Accordingly, $q \in \{1, \ldots, Q\}$ will be referred to as the {\it class index} and $s\in \{1, \ldots, S^q\}$ will be referred to as the {\it subclass index}. Indices $s=0$ and $s=S^q$ respectively represent an empty and full battery for vehicles with class index $q$. From this point on, we  use subscripts $s$ and superscripts $q$ to distinguish variables associated with cluster $(q,s)$.

With the help of this classifier, we can propose a scalable model of the scheduled EV load next.

\subsection{Interruptible Charging Modeled as Serial Subtasks}
We assume that the quantization points of the SoC, i.e., the $c_s^q$'s, are designed such that  it takes an EV with class index $q$ exactly one time step to get from SoC $c_s^q$ to $c_{s+1}^q$.
Each EV aims to reach the full charge  SoC $c_{S^q}^q$. Thus, charging can be modeled as a process in which the EV starts from an initial cluster $(q,s_{\rm init})$ and goes through a sequence of clusters to reach cluster $(q,S^q)$. For each class index $q$, these sequences of clusters are placed in tandem in a serial queueing system, i.e., EVs dispatched from the cluster $(q,s)$ are sent to $(q,s+1)$ and wait to get dispatched again. 
We assume that each time an EV moves from one cluster to the next, it needs to be authorized by the aggregator. This models charging as a number of {\it serially executed subtasks}, each of which are no longer interruptible. We denote the power consumption of EVs authorized to go from state $(q,s)$ to $(q,s+1)$ as $g_s^q,~s = 1, \ldots, S^q-1$ (see Fig. 1), highlighting the common {\it temporal charge evolution} of all EVs with class index $q$.

\begin{figure*}
\centering
\includegraphics[width = 0.7\linewidth]{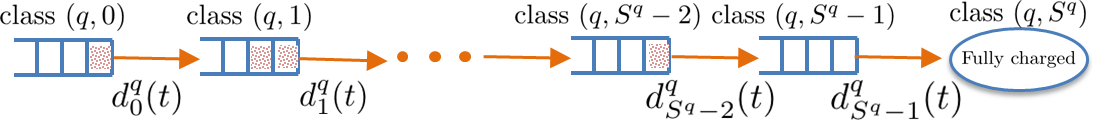}
\label{reg2}
\caption{Each charging task is divided up in a number of subtasks and appliances go through a serial queueing system to receive full charge}
\end{figure*}

\subsection{Decision Variables}
At each time epoch $t$, a population of $n_s^q(t)$ EVs belong to the cluster $(q,s)$. This number changes as appliances are authorized to charge and move forward in the subtask chain. We model these dynamics  by denoting the number of appliances authorized to move from cluster $(q,s)$ to $(q,s+1)$ {\it exactly at time $t$} as $d_s^q(t)$, and refer to it as the activation process. Alternatively, one can use the cumulative activation process $D_s^q(t)$,  which denotes the number of  appliances authorized to move from cluster $(q,s)$ to $(q,s+1)$ {\it at or before time $t$}, initialized as zero at $t = 0$. Consequently, for $t>0$, the aggregate charging demand of EVs is approximately equal to
\begin{equation}\label{lint}
L(t) \approx \sum_{q=1}^Q \sum_{s=1}^{S^q} d_s^q(t) g_s^q = \sum_{q=1}^Q \sum_{s=1}^{S^q} \left( D_s^q(t) - D_s^q(t-1) \right) g_s^q.
\end{equation}
Also, we can now write the dynamics of $n_s^q(t)$  as
\begin{eqnarray}\label{nclass}
n_s^q(t) &=& n_s^q(0) + \sum_{\ell=1}^{t-1} d_{s-1}^q(\ell) - \sum_{\ell=1}^{t-1} d_{s}^q(\ell) \nonumber\\ &=& n_s^q(0) + D_{s-1}^q(t-1) - D_s^q(t-1).
\end{eqnarray}

The values of $d_s^q(t)$ or $ D_s^q(t)$ are decision variables of our model, to be chosen by the scheduler (see Fig. 2).

\subsection{Ensuring Per Cluster Deadline Constraints} 
We denote by $m_s^q(t)$ the number of EVs that have a deadline to get activated from cluster $(q,s)$ at time $t$ {\it at the latest} or they would miss their global deadline.  Then, we define the {\it cumulative minimum activation process} for cluster $(q,s)$ as
\begin{equation}\label{bigM}
M_s^q(t) =   \sum_{\ell = 1}^t  m_s^q(\ell). 
\end{equation}
This provides a lower threshold on the number of activations from cluster $(q,s)$ {\it at or before time} $t$. To ensure hard QoS, the cumulative activation process of cluster $(q,s)$ should be bounded from below by the cumulative minimum activation process:
\begin{equation}\label{cum}
D_s^q(t) \geq M_s^q(t), ~~\forall t, s,q
\end{equation}

But, how should we determine the values of the $m_s^q(t)$'s?

\subsection{Mapping Ultimate Charge Deadlines to Subtask Deadlines}\label{subdead}
We denote that common global deadline shared by all EVs with class index $q$ as $k^q$.  Thus, all of these EVs should reach the cluster $(q,S^q)$ by $k^q$. In order to enforce this constraint, we need to map the global deadline of a charging task to individual subtask deadlines. We start by calculating $m_{S^q-1}^q(t)$, the number of appliances with a deadline to get activated from the cluster $(q,S^q-1)$ at time $t$ in order to finish charging and reach $(q,S^q)$ by time $k^q$. Since a transition from one cluster to the next takes exactly one step, $m_{S^q-1}^q(k^q-1)$ is equal to the total number of vehicles with class index $q$ that have an initial state of charge lower than or equal to $S^q-1$.  For all other time epochs $t \neq k^q-1$, $m_{S^q-1}^q(t) = 0$.

Similarly, we can derive $m_s^q(t)$, i.e. the number of EVs that have a deadline to get activated from cluster $(q,s)$ at time $t$. In order to reach cluster $(q,S^q)$ at or before time $k^q$, an appliance should have left cluster $(q,s)$ by time $k^q-(S^q-s)$. Thus, $m_s^q(k^q-(S^q-s))$ is equal to the number of EVs of type $q$ that have an initial SoC lower than or equal to $s$, and 0 for all other $t$'s.

 Note that the $m_s^q(t)$'s can be computed once and for all in an {\it offline} fashion before real-time operation. To do so, we initialize all $m_s^q(t)$'s at zero for all $t,q,s$. Then, for every appliance joining the program at initial cluster $(q,s_{\rm init})$, we update these terms as
\begin{equation}\nonumber
m_s^q(k^q - (S^q-s)) =  m_s^q(k^q - (S^q-s))  + 1, \forall s = s_{\rm init},\ldots,S^q-1.
\end{equation}
We can then calculate $M_s^q(t)$ from \eqref{bigM}.

\subsection{Decision Set}
With these definitions, we can now write the set of admissible values for the decisions variables $d_s^q(t)$:
\begin{align}\label{dcons2}
& d_s^q(t) \geq 0, \nonumber\\ &\sum_{\ell=1}^t d_s^q(\ell) \geq M_s^q(t), \nonumber\\
&  d_s^q(t) \leq n_s^q(t) = n_s^q(0) + \sum_{\ell=1}^{t-1} d_{s-1}^q(\ell) - \sum_{\ell=1}^{t-1} d_{s}^q(\ell), 
\end{align}
which are all affine. Alternatively, we can use the cumulative activation process $D_s^q(t)$ as the decision variable, with
\begin{align}\label{dcons3}
&  D_s^q(t) - D_s^q(t-1) \geq 0,\nonumber\\ & D_s^q(t) \geq M_s^q(t), \\
&  D_s^q(t) - D_s^q(t-1)  \leq n_s^q(0) + D_{s-1}^q(t-1) - D_s^q(t-1). \nonumber
\end{align}

The proposed load model has considerably reduced the size of the search space required to schedule EVs.
With this, we can now focus on scheduling the EVs in real-time.


\section{Scheduling Strategies}\label{sec.aggregator}
\subsection{Near-Optimal Solution: Model Predictive Scheduling}\label{mpc.sc}
 The aggregator's goal is to optimize $L(t)$ in order to minimize its cost for serving the load over the length of the night. Since the dispatch commands $a(t)$ are stochastic and their values are only revealed in a causal fashion, this is not a deterministic offline optimization and scheduling decisions should be made online as the  $a(t)$'s are received. 

Since current scheduling decisions affect future costs, an optimal schedule cannot be decided in a myopic fashion. Thus, one approach is to apply {\it model predictive control} (MPC) and optimize $L(t)$ such that the overall expected cost incurred in a look-ahead horizon is minimized \cite{bertsekas}. The cost incurred at different intervals in the look-ahead horizon should be modeled separately. Following the aggregator's objective explained in Section \ref{aggobj}, we use a function $C^{\rm Penalty}(.)$ to capture the intra-hourly penalties paid if deviating from $P^{(k)}(t)+ a(t)$, and a function $C^{\rm Market}(.)$ to capture the costs of updating the bulk purchase from the market to ensure hard QoS, cf. \eqref{agg.opt}. Here, the scheduler needs to solve an integer program to find the optimum values of $D_s^q(t)$ for all $t$ in the look-ahead horizon. 
By making dummy decisions about future $D_q^q(t)$'s and $L(t)$, the MPC helps find a suitable update for the future bulk purchase in the hourly electricity market, i.e., it helps optimize $P^{(k+1)}(t)$. We will showcase this important aspect in our numerical experiments. Due to lack of space, we defer the thorough analysis of the MPC approach  to future work. Instead, the question that we focus on in this paper is as follows: Can we find a simple {\it heuristic policy} that has similar performance to MPC, with lower computational complexity?

\subsection{Proposed Heuristic: SPUC}\label{heur}

The heuristic policy we propose is non-anticipative and tries to minimize the current deviation penalty $C^{\rm Penalty}(.)$, without considering future penalties and market costs. Thus, it does not carry out the adaptive updates on the bulk purchase, $P^{(k)}(t)$, and simply follows $P^{(0)}(t) + a(t)$, with $P^{(0)}(t)$ determined at $t=0$.
 Thus, at each  $t$, the scheduler activates a number of subtasks so as to to get the consumption as close as possible to $P^{0}(t) + a(t)$. But which subtasks should be picked? If any subtask has reached its deadline, the scheduler has no choice but to activate it. Thus, the first step would be to activate all the subtasks with zero laxity. If the consumption has not exceeded $P^{(0)}(t) + a(t)$ yet, the scheduler then picks a number of subtasks to activate among those remaining. A heuristic algorithm needs a metric to rank the waiting subtasks from different clusters $(q,s)$ in the order of  time sensitiveness. Our proposed metric, referred to as {\it Slack per Unit Charge} (SPUC), is used by the following algorithm at every $t$:
\begin{enumerate}
\item Schedule subtasks with immediate deadlines: $\forall q,s$, 

If $D_s^q(t) < M_s^q(t)$: Set $d_s^q(t)  = M_s^q(t) - D_s^q(t)$,

Else: Set $d_s^q(t)  = 0$;
\item Set $L(t) = \sum_{q=1}^Q \sum_{s=1}^{S^q} d_s^q(t) g_s^q $;
\item Define $\mathcal{S}$ as the set of all clusters $(q,s)$; 
\item While $L(t) \leq P^{(0)}(t) + a(t)$ and $\mathcal{S} \neq \emptyset$
\begin{enumerate}
\item Pick the cluster $(q^\dagger,s^\dagger)$ in $\mathcal{S}$  with the largest value for the SPUC metric $\chi^q_s(t)$ (defined later);
\item If $d^{q^\dagger}_{s^\dagger}(t) = n^{q^\dagger}_{s^\dagger}(t)$ $\rightarrow$ no EV left to schedule. Remove  $(q^\dagger,s^\dagger)$  from $\mathcal{S}$ and go back to step (a);
\item Increase $d^{q^\dagger}_{s^\dagger}(t)$ and $D^{q^\dagger}_{s^\dagger}(t)$ by 1;
\item Set $L(t) = L(t) + g ^{q^\dagger}_{s^\dagger}$;
\end{enumerate}
\item Activate $d_s^q(t)$ subtasks in each cluster $(q,s)$;
\end{enumerate}

But how to design the metric $\chi^q_s(t)$? The scheduler can potentially apply an Earliest Deadline First/ Least Laxity First (LLF) policy among all clusters. The slack of any subtask waiting in cluster $(q,s)$ is simply equal to
\begin{align}\label{agg.opt}
\rho_s^q(t) =  k^q - ( t + (S^q - s) )
\end{align}
Thus, an LLF policy would find the cluster with the lowest $\rho_s^q(t)$, and activate one task from that cluster. This process continues until $L(t) $ exceeds $P^{(0)}(t) + a(t)$.

 However, the slack for each subtask (accounted for by $\rho_s^q(t)$) is shared with the subsequent subtasks that compose the global charge task. Thus, if all the slack is used up for a specific subtask, the remaining subtasks in the chain need to be activated immediately with zero flexibility. Thus,  a more appropriate benchmark is to normalize $\rho_s^q(t)$ by the number of remaining subtasks for appliances in cluster $(q,s)$ to reach $(q,S^q)$, which is equal to $S^q-s$. This gives us our initial per cluster time-sensitivity index, which we denote by $\xi^q_s(t)$:
\begin{align}
& \xi^q_s(t) = \frac{\rho_s^q(t)}{S^q-s}.
\end{align}

Using $\xi^q_s(t)$ does not distinguish between subtasks with higher and lower power consumption. An EV using a level-3 high power charger can prove more problematic in terms of fitting in a capacity constrained system and should be prioritized. Consequently, we normalize $\xi^q_s(t)$ by the average charge power for the remaining subtasks after state $s$, i.e., by $$\bar{g}_s^q = \frac{\sum_{h=s}^{S^q-1} g_h^q}{S^q-s}.$$
This gives us our final ranking benchmark for sorting the time sensitivity of scheduling a subtask from cluster $(q,s)$:
\begin{align}
& \chi^q_s(t) = \frac{\rho_s^q(t)}{\bar{g}_s^q(S^q-s)},
\end{align}
referred to as the SPUC metric. Ties are broken by comparing the variance of charge power for the {\it remaining} subtasks in the chain after cluster $(q,s)$, i.e., by finding $(q,s)$ with the highest
\begin{align}\mbox{Var}({g})_s^q = \sum_{h=s}^{S^q-1} (g_h^q-\bar{g}_s^q)^2,\end{align}
which is a measure of how {\it bursty} the remaining charge profile for a certain appliance is.
By reapplying this criterion in a loop, we can pick the appropriate $d_s^q(t)$'s for each cluster.

\begin{remark} The performance of both schedulers can be assessed under the following effects: 1) users may depart earlier than their declared deadline; 2)  real-time arrivals modeled by an {\it arrival process} can be considered instead of task being submitted a-priori; 3) classification of tasks may be overly coarse; or 4) users may provide inexact cluster estimates.  Papers considering the last issue include \cite{w2} \cite{gupta2}. \end{remark}

\section{Numerical Experiment}
To assess the performance of the proposed scheduling methods in Sections \ref{mpc.sc} and \ref{heur}, we compare their real-time performance under a realistic setting. We simulate the operations of an aggregator in charge of directly scheduling an average population of 1000 EVs over a 12 hour nighttime period between 9 pm to 9 am. Twenty different characteristic clusters $q$ were considered, bundled into 4 groups. Each group shares a certain shape of the charging pulse $g_1^q,\ldots,g_{S^q}^q$: a rectangular pulse with a height of 1.1 kWs and a maximum length of 8 hours; A triangular pulse with a height of 2.2 kWs and a maximum length of 8 hours;  A rectangular pulse with a height of 3.3 kWs and a maximum length of 4 hours; A triangular pulse with a height of 6.6 kWs and a maximum length of 4 hours. The initial SoC was drawn from lognormal distributions consistent with the findings in \cite{evpapersg} based on actual PHEV charging data. We used $\ln \mathcal{N}(3,1.2^2)$ for the first two cluster and $\ln \mathcal{N}(1,0.58^2)$ for the next two. Clusters bundled in the same group differ in terms of the quantized deadline $k^q$. We generated the hard deadline for each request randomly according to the real-world based statistics in \cite{evpapersg}, and then quantized it into one of the 5 possible options. The wind following signals were generated using real-world wind forecast errors from the Bonneville Power Administration with a 5 minute resolution, normalized to a maximum wind following capacity of 60 kWs for the aggregator. Wind following  is only provided in the first 9 hours of the 12 hour night period.

To solve the integer program for the MPC scheduler, we used CVX + Gurobi \cite{cvx}. A linear cost was assumed, with the intra-hour deviation penalties assumed to be 10 times higher than energy market costs. If each hour is $\Delta$ units, at time $t$ (within hour $k$), the cost at time $\ell>t$ in the look-ahead is:
\begin{align}\label{agg.opt}
C^{\rm Penalty}(\ell) &= 10\left\|  L(\ell) - P^{(k)}(\ell) - \mathbb{E}[a(\ell)]  \right\|_1, t \leq \ell \leq k\Delta\nonumber\\ C^{\rm Market}(\ell) &= \| L(\ell) - P^{(k)}(\ell)\|_1, \ell \geq k\Delta
\end{align}
A persistence time series model was applied to forecast expected future intra-hour values of $a(t)$, while future inter-hour values were predicted with their marginal expected value, i.e., 0. The average run-time for one full night period (144 epochs) exceeds one hour on a 2.67 GHz i7 CPU. Notice that the last update on the bulk purchase performed by the MPC  is $P^{(11)}(t)$, since the length of the night is 12 hours and the last update can happen at hour $k=11$. Fig. 3 displays $P^{(0)}(t)$, $P^{(11)}(t) + a(t)$, and the consumption of the aggregator under the MPC scheduler, $L^{\rm mpc}(t)$, following $P^{(k)}(t) + a(t)$. The total deviation of $L^{\rm mpc}(t)$ from the dispatch $P^{(k)}(t) + a(t)$ is nearly zero (0.03 kWhs over the length of the night).

\begin{figure}
\centering
\includegraphics[width = 0.82\linewidth]{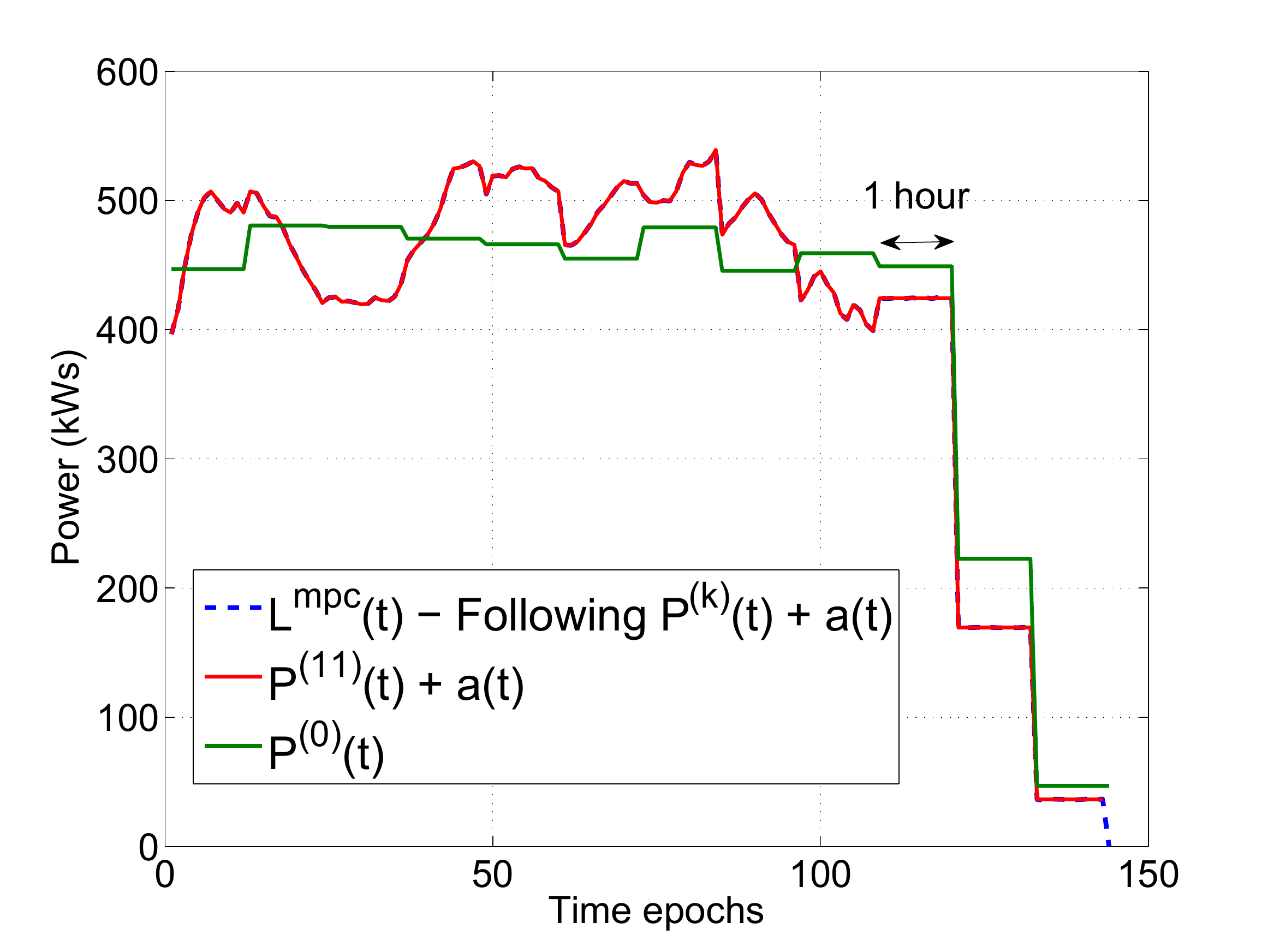}
\label{sim1}
\caption{Performance of the MPC scheduling method}
\end{figure}
\begin{figure}
\centering
\includegraphics[width =0.79\linewidth]{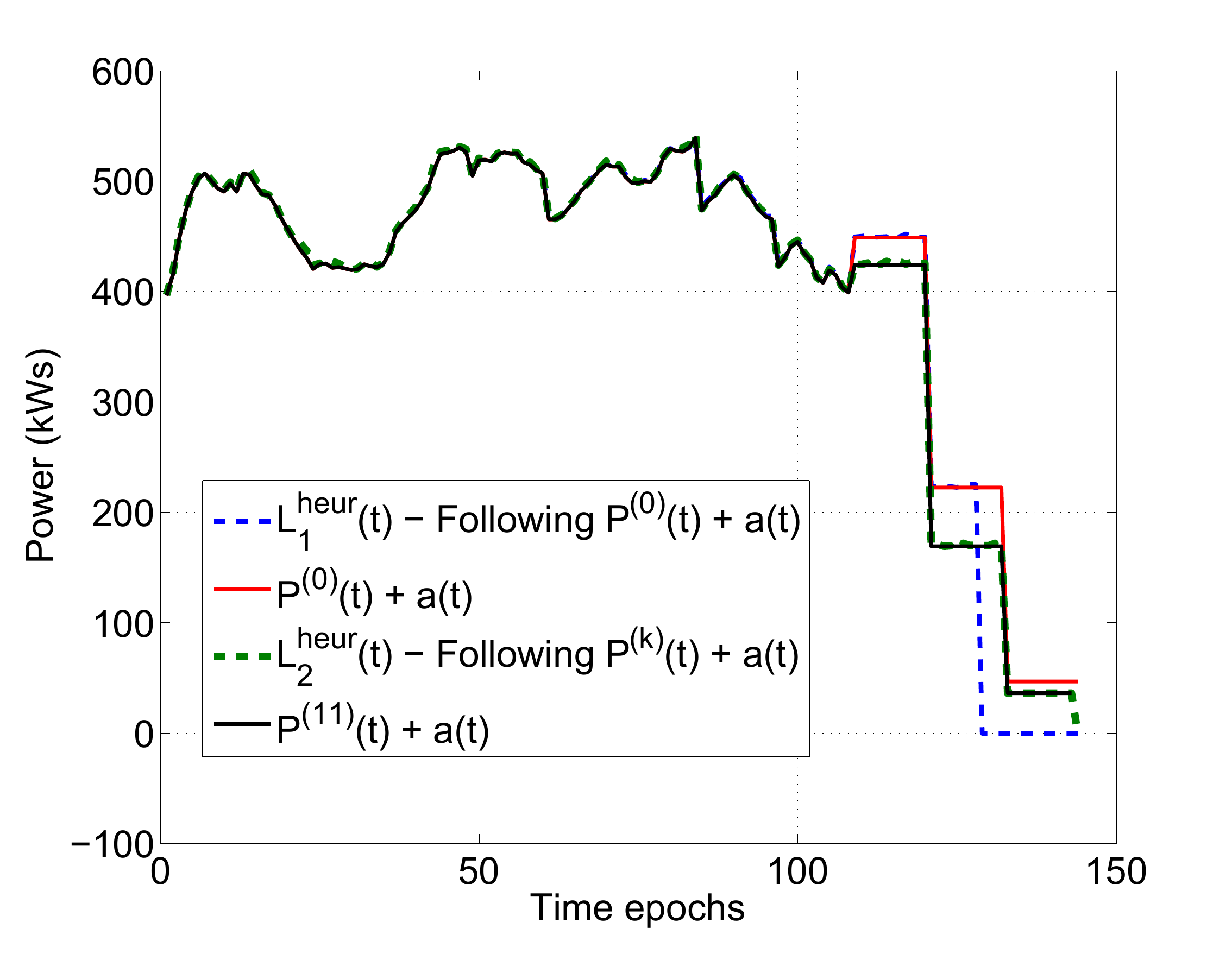}
\label{sim2}
\caption{Performance of the heuristic scheduling method}
\end{figure}

The heuristic method delivers a comparable performance to that of the MPC, with a run-time close to one minute. The increase in deviation from ISO dispatch, though considerable, is still negligible (13 kWhs overall). However, remember that the heuristic method itself does not carry out any hourly updates on the bulk purchase. We simulate the performance of the heuristic technique when following  $P^{(0)}(t) + a(t)$.  Notice the poor performance in the last 2 hours ($L^{\rm heur}_1(t)$).
However, if presented with the adaptively updated bulk purchase,  the heuristic method successfully follows the pseudo optimal profile $P^{(k)}(t) + a(t)$, as seen in Fig. 4  ($L^{\rm heur}_2(t)$).

\section{Acknowledgments}
This work was supported by NSF grants CNS-1228717, CCF-1229008, and the U.S. DOE through the Consortium for Electric Reliability Technology
Solutions (CERTS).

\bibliographystyle{IEEEtran}
\small
\bibliography{New2}

\end{document}